\documentclass[aps,prl,twocolumn,groupedaddress]{revtex4}

\usepackage{graphicx}	
\usepackage{dcolumn}	
\usepackage{bm}		

\begin{document}


\title{Compton Scattered Transition Radiation from Very High Energy Particles}


\author{Michael L.\ Cherry}
\email{cherry@lsu.edu}
\author{Gary L.\ Case}
\affiliation{Dept.\ of Physics and Astronomy, Louisiana State 
      University, Baton Rouge, LA 70803}

\date{\today}

\begin{abstract}
X-ray transition radiation can be used to measure the Lorentz factor of relativistic 
particles. At energies approaching $\gamma = E/mc^2 = 10^5$, transition radiation 
detectors (TRDs) can be optimized  by using thick ($\sim 5 - 10$ mil) foils with 
large ($\sim 5-10$ mm) spacings. This implies X-ray energies $\agt 100$ keV and the 
use of scintillators as the X-ray detectors. Compton scattering of the  X-rays out of 
the particle beam then becomes an important effect. We discuss the design of very high energy detectors, the use of metal radiator foils rather than 
the standard plastic foils, inorganic scintillators for detecting Compton scattered transition radiation,
and the application to the ACCESS cosmic ray experiment.
\end{abstract}


\maketitle

%

\section{I.\ Introduction}

Transition radiation, originally predicted by Ginzburg and Frank in 1946 \cite{Ginzburg}, has been used as the basis of detectors of high energy particles for over thirty years \cite{TerMikaelian, Favuzzi}. Because the detectors are typically thin in terms of g/cm$^2$, the incident particle passes 
\begin{figure}[b]
\includegraphics[width=240pt]{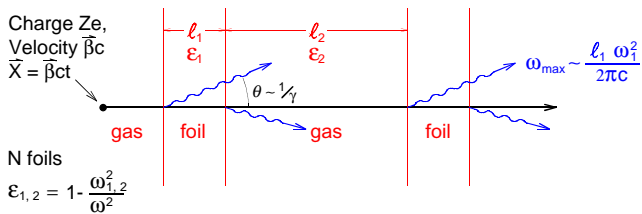}
\caption{Transition radiation geometry, with particle at position $\vec{x} = \vec{\beta}ct$ moving 
 from left to right crossing the boundary from foil (with thickness $l_1$) to gas (with a distance $l_2$ between the foils) at $t=0$. \label{TRschematic}}
\end{figure}
through the detector without interacting and at essentially constant velocity. In particle physics or high energy astrophysics experiments where 
an incident particle must be identified without being destroyed, or in space applications where weight is precious and a calorimeter would be prohibitively massive, transition radiation detectors have been shown to be effective and useful devices \cite{Favuzzi, CherryBari}. Both at 
accelerators and in space, a common application involves particle identification 
(e.g., lepton/hadron discrimination) at fixed energy: Since the X-ray transition 
radiation intensity produced by a charged particle crossing a single interface 
between two different materials increases linearly with increasing particle Lorentz factor, the particle mass can be determined by combining independent energy knowledge with the TRD's 
information about Lorentz factor, typically at Lorentz factors up to 
$\gamma \sim 10^3 - 10^4$. A more challenging application is the measurement of the particle energy:
In an experiment on the Space Shuttle \cite{CRN1,CRN2}, the 
energy of cosmic ray nuclei was determined using the linear increase of the transition X-ray signal 
with $\gamma$, again at Lorentz factors $\gamma \sim 10^3.$ At higher Lorentz factors, the yield from
standard transition radiation detectors saturates. In order to extend the applicability of the technique 
to the accurate measurement of energies in the regime near $\gamma \sim 10^5$ (as is required, for example, for NASA's proposed ACCESS cosmic ray experiment 
\cite{ACCESS1,ACCESS2,Case,Wakely}), one must modify the standard 
approaches to detecting transition radiation. We discuss here the possibility of optimizing detectors for high energy particles by using hard X-rays (above $\sim$ 100 keV) Compton scattered out of the incident particle beam.

\section{II.\ Transition Radiation Detectors FOR Very HIGH ENERGies}

As a charged particle crosses the interface between two media, it must rearrange its electromagnetic 
fields in order to satisfy Maxwell's Equations and their boundary conditions at the interface. The radiation 
intensity can be derived from the Li\'{e}nard-Wiechert potentials \cite{Cherry78}:
\begin{equation}
\frac{d^2S_o}{d\Omega d\omega} = \frac{1}{c} {\left( \frac{Ze \omega \beta \sin{\theta}}{2 \pi} \right)}^2 
   {\left| \int_{-\infty}^{\infty} \epsilon^{1/4} e^{i \omega t (1-\epsilon^{1/2}\beta \cos{\theta})} dt 
   \right|}^2 \hspace{2pt} .
\label{Lienard}
\end{equation}
\noindent Here $d^2S_o/d\Omega d\omega$ is the energy emitted per unit solid angle $\Omega$ and X-ray 
frequency $\omega$, $Ze$ is the charge of the incident particle, $\beta = v/c$ is its velocity, and $\theta$ 
is the angle of the X-ray photon with respect to the particle's trajectory. The integral is performed 
from time $ t = -\infty$ to $t = \infty$, with the particle crossing the interface at 
$t=0$. The dielectric ``constant"  is $\epsilon_1$ to the left of the interface (corresponding to $ t < 0$) and 
$\epsilon_2$ to the right (where $t > 0$), with $\epsilon_{1,2} = 1 - \omega^2_{1,2}/\omega^2$, where 
$\omega_{1,2}$ is the plasma frequency in material 1 or material 2. For definiteness, we take 
$\omega_1 \gg \omega_2$ in the following (Fig.\ \ref{TRschematic}).
 
The largest contribution to the integral is from times 
\begin{equation}
\tau_{1,2} \leq \frac{1}{\omega ( 1-\epsilon_{1,2}^{1/2}\beta \cos{\theta})}
\end{equation}
and pathlengths
\begin{equation}
\beta c \tau_{1,2} = \frac{Z_{1,2}}{2} = 
    \frac{2 \beta c}{\omega} \left(\frac{1}{\gamma^2}+\frac{\omega^2_{1,2}}{\omega^2}+\theta^2 \right)^{-1}  .
\end{equation}
The length $Z_{1,2}$ here is referred to as the formation zone in medium 1 or 2 respectively, 
where we use the approximations that $\gamma \gg 1$, 
$\omega_{1,2} \ll \omega$, and $\theta \ll 1$. Physically, the formation zone describes the longitudinal
extent of the particle's field along the beam direction. As long as $\omega \alt \gamma \omega_1$, the 
field extends farther forward ahead of the particle in the less dense material (e.g., gas or vacuum) 
than in the denser (e.g., solid foil) material; i.e., $Z_2 > Z_1$. At high frequencies, however, the 
formation zone length becomes independent of the material:
\begin{equation}
Z_{1,2} \longrightarrow \frac{4 \beta c}{\omega} \left(\frac{1}{\gamma^2}+\theta^2 \right)^{-1} 
\hspace{6pt} {\rm for} \hspace{12pt} \omega \gg \gamma \omega_1 \gg \gamma \omega_2  \hspace{12pt} .
\end{equation}

For a single interface, the integration in Eq.\ \ref{Lienard} gives
\begin{equation}
\frac{d^2S_o}{d\Omega d\omega}=\frac{1}{c} {\left( \frac{Ze \omega \theta}{4 \pi c} \right)}^2  
     \left( Z_1-Z_2\right)^2  \hspace{6pt} .
\label{SingleInterface}
\end{equation}
The intensity depends on the difference between the formation zone lengths.
One can picture the particle decelerating from $\beta$ to 
0 as it leaves material 1, and then immediately accelerating back to its original velocity as it enters 
material 2. The effect is then closely related to that of bremsstrahlung: The net 
intensity as the particle crosses the interface is the square of the summed amplitude of the bremsstrahlung 
in medium 1 and medium 2, where the two amplitudes differ in phase by $\pi/2$. In the case of two identical 
materials, the effect disappears. 

\begin{figure}
\includegraphics[trim = 20 0 25 0, width = 260pt]{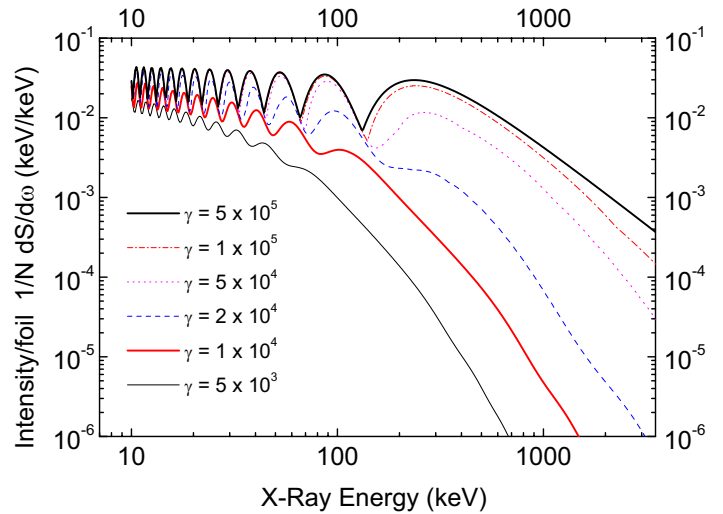}
\caption{Calculated transition radiation X-ray spectra for particle Lorentz factors ranging from  $\gamma = 5 \times 10^3$ to $5 \times 10^5$. The radiator consists of 150 Teflon foils with thickness  $l_1 = 370\  \mu$m and vacuum gaps $l_2 = 1$ cm. The intensity is calculated for the case of no absorption (i.e., $\mu = 0$). \label{calculated_spectra}}
\end{figure}

At relatively low X-ray energies, the spectrum 
(\ref{SingleInterface}) varies only slowly with $\omega$. At frequencies 
$\omega \gg \gamma \omega_1 \gg \gamma \omega_2$, however, the bremsstrahlung amplitudes cancel (i.e., $Z_1 - Z_2 \rightarrow 0$) and the frequency spectrum cuts off. As $\gamma$ increases, the spectrum extends to higher frequencies linearly with $\gamma$, and the energy dependence of the emitted intensity arises from the extension of the spectrum to harder frequencies as 
the particle energy grows. For high particle energies, the cutoff in the spectrum occurs at correspondingly high X-ray energies. Integrating Eq.\ \ref{SingleInterface} gives the total yield from a single interface:
\begin{equation}
S_o = \frac{Z^2}{3} \alpha \hbar \frac{(\omega_1-\omega_2)^2}{\omega_1+\omega_2} \gamma 
     \sim \frac{Z^2}{3} \alpha \hbar \omega_1 \gamma \hspace{12pt} ,
\end{equation}
where $\alpha = e^2/\hbar c$ is the fine structure constant. For a typical plasma frequency $\omega_1 \sim 20 - 30$ eV and $\gamma = 10^5$, $\gamma \omega_1  \sim 2 - 3 $ MeV and the total intensity produced by a singly charged particle in a radiator with N = 100 foils is on the order of an MeV.  

Near the upper end of the spectrum (i.e., near $\omega \sim \gamma\omega_1$),  
the characteristic number of photons emitted by a singly charged particle at a single interface is small ($S_o/\hbar \omega \sim \alpha/3$).
Therefore, one typically sums the signal from a stack of a hundred or 
more interfaces provided by foils with thickness $l_1$ and plasma frequency $\omega_1$ spaced by gaps 
with corresponding values $l_2$ and $\omega_2$ (or an equivalent foam or fiber geometry). Absorption can 
be included by allowing the wave vector $\vec{k}$ to be complex. 
If the X-ray absorption cross section is taken to be $\mu = Im(k)$, then the intensity from a stack of 
N foils can be computed from the coherent sum of the amplitudes from the individual 
interfaces \cite{Cherry78,Yodh}:
\begin{widetext}
\begin{equation}
\label{coherentsum}
\frac{d^2S_N}{d\Omega d\omega} = \frac{d^2S_o}{d\Omega d\omega} e^{-(N-1)(\mu_1l_1+\mu_2l_2)/2} \times 
 4\sin^2{(l_1/Z_1)} \frac{\sinh^2{\frac{N}{4}(\mu_1l_1+\mu_2l_2)}+\sin^2{N(l_1/Z_1+l_2/Z_2)}}
{\sinh^2{\frac{1}{4}(\mu_1l_1+\mu_2l_2)}+\sin^2{(l_1/Z_1+l_2/Z_2)}} \hspace{12pt} . 
\end{equation}
\end{widetext}

The result of integrating Eq.\ \ref{coherentsum} over angles is shown in Fig.\ \ref{calculated_spectra},
where the interference effects due to the coherent sum over multiple boundaries 
result in a modulation of the multi-foil spectrum around the single foil spectrum. 
The mild energy dependence of the yield at low X-ray frequencies and the energy-dependent cutoff 
$\gamma \omega_1$ at high frequencies are clearly visible here.
In the case of Lorentz factors $\sim 10^3 - 10^4$ (i.e., tens of keV X-ray energies) \cite{Cherry78}, 
1) the last (highest frequency) interference maximum in the spectrum occurs near 
\begin{equation}
\label{omega_max}
\omega_{\rm max} = \frac{l_1 \omega_1^2}{2 \pi c} \hspace{12pt} ;
\end{equation}
and 2) the integrated yield increases linearly with energy up to a 
saturation Lorentz factor given by
\begin{equation}
\label{gamma_s}
\gamma_s \sim \frac{0.6 \omega_1}{c} \sqrt{l_1l_2} \hspace{12pt} .
\end{equation}
As $\gamma$ increases, additional high-frequency peaks appear in the spectrum until, near $\gamma_s$, the
final maximum appears at $\omega_{\rm max}$ (Fig.\ \ref{calculated_spectra}). Increases of $\gamma$ beyond 
$\gamma_s$ result in little additional increase in the multiple foil 
interference maxima even though the single interface yield $S_o$
continues to grow as it extends to higher frequencies with increasing Lorentz factor. 

The expression for $\gamma_s$ depends on the implicit assumption that the detector is 
tuned to be sensitive near $\omega_{\rm max}$. The energy dependence is 
largest near the cutoff $\gamma \omega_1$, though, and at sufficiently high Lorentz factors, $\gamma \omega_1$ 
can be large compared to $\omega_{\rm max}$. 

The prescription for building a TRD sensitive to high particle 
Lorentz factors (e.g., $\gamma_s \sim 10^5$) is therefore as follows:\\
$\bullet$ First, according to Eq.\ \ref{gamma_s}, one must increase the dimensions $l_1$ and $l_2$ and 
increase the foil 
density (i.e., increase $\omega_1$). The integrated yield at saturation is then $S_N \propto
Z^2 N \omega_1 \gamma_s$ and the number of photons produced is $N_{\gamma} \propto S_N / \omega_{\rm max} 
\propto N \gamma_s/l_1 \omega_1$. For a space experiment in which height (i.e., distance along the beam) is
limited, then (assuming $l_2 \gg l_1$) the total length of the detector is $L \sim Nl_2$ and the number of 
photons produced becomes 
\begin{equation}
N_{\gamma} \propto \frac{L}{l_2} \frac{\gamma_s}{l_1 \omega_1} \propto \frac{\omega_1}{\gamma_s} \hspace{12pt} .
\end{equation}
As $\gamma_s$ goes up, the number of photons goes down unless one compensates by increasing $\omega_1$. 
Increasing the dimensions and density means that the characteristic frequencies increase 
(Eq.\ \ref{omega_max}). \\
$\bullet$ Second, as the frequency spectrum becomes harder, the detector must become thicker. As X-ray 
energies increase above 100 keV, gas detectors become inefficient and scintillators become more suitable X-ray 
detectors. 	\\
$\bullet$ Third, as the spectrum becomes harder, the probability of Compton scattering grows. This results 
in the X-ray signal (which is initially emitted in the forward direction) being scattered away from the particle 
trajectory, so that the X-rays and the ionization become spatially separated. Large area 
segmented detectors can then be used to detect X-rays away from the incoming particle path.

\section{III.\ Radiation from Metal Foils}

For a typical TRD, low-density plastic foils or foam are used in order to minimize the X-ray absorption. For 
the case of very high energies, though, where it is desirable to increase $\omega_1$, metal foils may be 
the best choice. In this case, the X-rays are sufficiently hard that photoelectric absorption is not a major 
concern, and the increased radiator grammage can be used to advantage by increasing the Compton scattering probability. 

In the standard case of an insulating radiator, the X-ray wave vector is 
$k = \sqrt{\epsilon_{1,2}} \omega/c = \omega/c (1 - \omega_{1,2}^2/2 \omega^2)$ and 
the relevant phase factor is
\begin{equation}
i (\vec{k} \cdot \vec{x} - \omega t) = - \frac{2i \beta c t}{Z_{1,2}}
\end{equation} 
for $k$ real. For a metal, the wave vector depends on the conductivity and is complex:
\begin{equation}
k^2 = \mu_{1,2} \epsilon_{1,2} \frac{\omega^2}{c^2} 
\left( 1 + i \frac{4 \pi \sigma}{\omega \epsilon_{1,2}} \right) \hspace{12pt} ,
\end{equation}
where $\mu_{1,2}$ is the permeability \cite{Jackson}. Here 
\begin{equation}
\sigma = \frac{\omega_{1,2}^2 \rho}{4 \pi (g - i \omega)} 
   \hspace{12pt} \rm{where} \hspace{12pt}
   \rho = \left\{ \begin{array}{ll} 0 & \rm{for\  an\  insulator}\\ 
                                    1 & \rm{for\  a\  metal} \end{array}  \right.   
\end{equation}
and $g = \omega_{1,2}^2 \rho / 4 \pi \sigma_o$ is a constant related to the conductivity $\sigma_o$ at 
zero frequency.  For a good nonmagnetic ($\mu_{1,2} = 1$) conductor at high frequencies 
($\omega \gg \omega_{1,2} \gg g$), 
\begin{equation}
k = \frac{\omega}{c} 
\left(1-\frac{\omega_{1,2}^2}{\omega^2} + \frac{i}{2} \frac{\omega_{1,2}^2}{\omega^2}\frac{g}{\omega}\right)
  \hspace{12pt} .
\end{equation}
The transition radiation intensity is again given by 
Eq.\ \ref{coherentsum}, where now the effective foil plasma frequency is $\omega_1 \sqrt{1 + \rho}$ and 
the effective absorption cross section is 
\begin{equation}
\mu = Im(\vec{k}) =
{\left( \frac{\omega_1^2}{2 \omega^2} \right)}^2 \frac{\rho \omega^2}{2 \pi \sigma_o c} \hspace{12pt} .
\label{mu}
\end{equation}

In the standard case, photoelectric absorption in the plastic foils suppresses the intensity at low frequencies. For the case of aluminum foils at 100 keV, Eq. \ref{mu} gives an effective conductivity-induced cross section value of $\mu \sim 2 \times 10^{-4}$ cm$^{-1}$ and the photoelectric cross section is $\sim 2 \times 10^{-2}$ cm$^{-1}$, so that there is comparatively little suppression due to either the conductivity component or the photoelectric component of $\mu$ in the metal at these energies. The characteristic frequency and saturation Lorentz factor become
\begin{equation}
\omega_{\rm max}' = (1 + \rho) \frac{l_1 \omega_1^2}{2 \pi c} \hspace{12pt} 
\end{equation}
and
\begin{equation}
\gamma_s' \sim \frac{0.6 \omega_1}{c} \sqrt{(1 + \rho) l_1l_2} \hspace{12pt} .
\end{equation}
Metal foils cause the yield to increase (because the intensity increases with density and $\omega_1$ increases to $\sqrt{2} \omega_1$) and to extend to higher X-ray frequencies. In the general case of a conducting and a nonconducting foil radiator with the same density, $\gamma_s$, thickness (in g/cm$^2$), and total length $N(l_1+l_2)$, the total produced intensity from the conducting foil radiator is twice that produced from the nonconductor. 

In Fig.\ \ref{calculated_spectra}, the spectra shown for electrons passing through a Teflon foil radiator are calculated from Eq.\ \ref{coherentsum} integrated over angles with $\mu = 0$. Fig.\ \ref{comparison} shows spectra calculated for $Z$ = 1 particles at $\gamma = 10^5$ in aluminum and Teflon foil radiators with the same $\gamma_s$ and $\omega_{\rm max}$, and $\mu$ given by Eq.\ \ref{mu}. (Teflon is chosen for the comparison because of its high density $\rho = 2.2$ g/cm$^2$ and correspondingly high plasma frequency.) In Fig.\ \ref{comparison}, N is the same for the two radiators but the material traversed by the particle in passing through the Teflon radiator is twice that traversed through the aluminum radiator. The yield per foil for aluminum is then approximately 1.3 times the yield for Teflon above 100 keV and 1.2 times higher than for Teflon above 30 keV. The figure also shows the effect for the aluminum radiator of Compton scattering the photons out of the particle beam and detecting them in a scintillator, including the effects of photoelectric absorption, fluorescence, and photoelectron statistics. Photons are assumed to be produced at random positions moving forward along the particle track, and then absorbed and/or Compton scattered through the residual detector material by a Monte Carlo photon transport code. The detector geometry becomes important in this case: We assume the geometry of the Compton Scatter Transition Radiation Detector (CSTRD) proposed for ACCESS (cf.\ Sec.\ IV and Fig.\ \ref{ACCESS} below). Photons are counted as Compton scattered if they are separated from the incident particle beam by a minimum of 1.9 cm.

\section{IV.\ Compton Scattered Yield vs.\ Energy}

A xenon-filled wire chamber sensitive to tens of keV may be suitable for a TRD \begin{figure}
\includegraphics[trim = 20 0 25 0, width = 260pt]{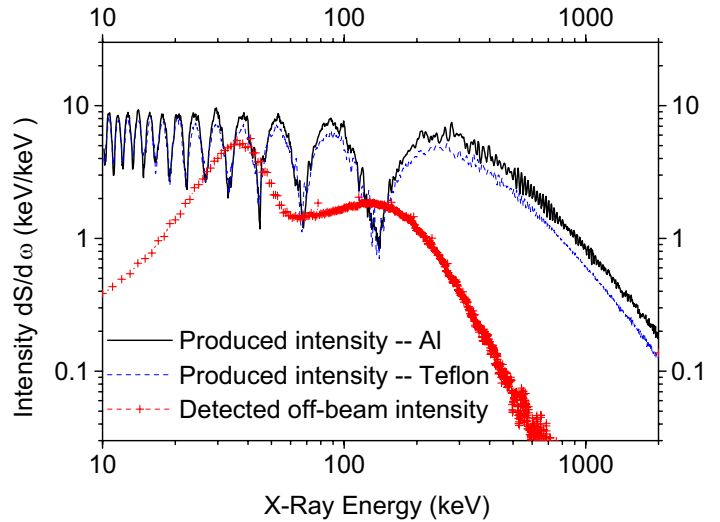}
\caption{Calculated transition radiation \underline{produced} by protons with $\gamma = 10^5$ in aluminum/vacuum (N = 150, $l_1 = 150\  \mu$m, $l_2 = 1$ cm, $\gamma_s = 1.7 \times 10^5,\  \omega_{\rm max} = 260$ keV, upper curve) and Teflon/vacuum (N = 150, $l_1 = 370\  \mu$m, $l_2 = 1$ cm, $\gamma_s = 1.7 \times 10^5,\  \omega_{\rm max} = 260$ keV, middle curve) radiators. Bottom curve shows a calculation of the \underline{detected} X-ray energy in detector elements away from the particle beam, taking into account photoelectric absorption, Compton scattering, fluorescence, detector efficiencies, and resolution for the ACCESS detector design shown in Fig.\ \ref{ACCESS}. The bottom curve of detected intensity is calculated for noninteracting particles passing vertically down through the center of the detector. \label{comparison}}
\end{figure}
at Lorentz factors $\sim 10^3 - 10^4$, but when $\gamma \omega_1$ exceeds $\sim 100$ keV energies, thicker detectors are needed (e.g., scintillators) in order to obtain the maximum sensitivity to particle energy. 
\begin{figure*}
\includegraphics[trim = 25 0 25 0, width = 300pt]{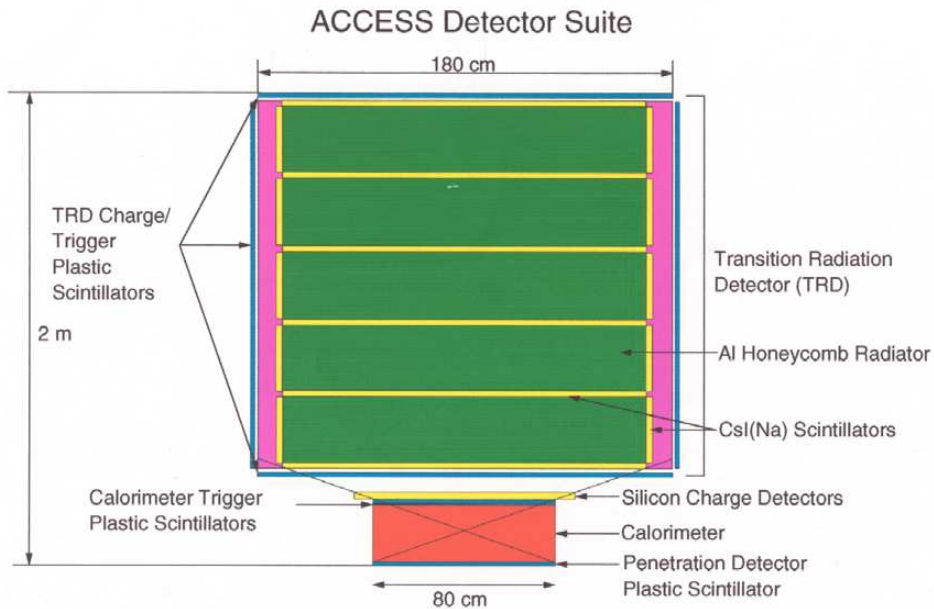}
\caption{Schematic showing the Compton Scatter Transition Radiation Detector for
the proposed ACCESS high energy cosmic ray composition experiment\cite{mitchell}. \label{ACCESS}}
\end{figure*}

In the standard tens of keV case, the low frequency part of the spectrum is attenuated by photoelectric absorption in the radiator and high energy photons
escape from the thin gas-filled detector  \cite{Cherry78}. Nevertheless, the
detector is typically kept thin because the radiation is mainly emitted forward along the particle trajectory. 
\begin{figure}[b]
\includegraphics[trim = 0 0 25 0, width = 260pt]{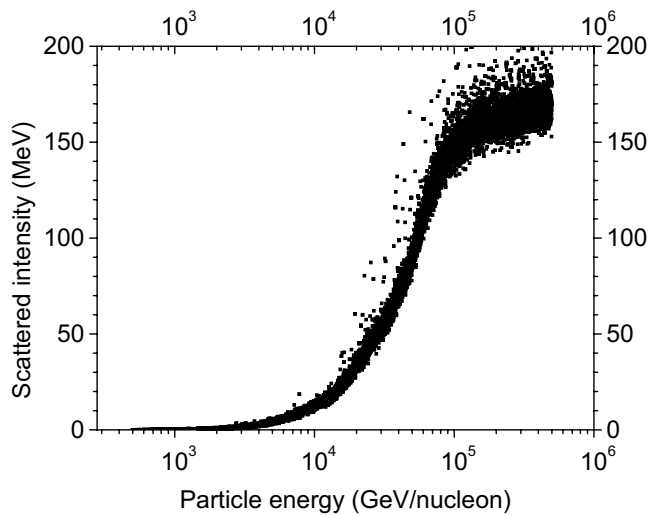}
\caption{Monte Carlo simulation of the transition radiation yield for iron nuclei in a detector configuration suitable for measuring the energies of high energy cosmic ray nuclei \cite{Case}. The detector parameters are listed in the text. The vertical axis gives the intensity due to Compton scattered X-ray ``hits'' detected away from the particle trajectory. \label{yield_vs_gamma}}
\end{figure}
The X-rays are then detected in the presence of the particle's ionization, and the ratio of X-ray ``signal'' to ionization ``background'' is maximized in a thin detector. In the high energy case (i.e., above $\sim$ 100 keV), Compton scattering becomes more important than photoelectric absorption. The produced X-rays are scattered away from the particle trajectory, and the signal can be measured in the absence of the ionization. One can then use a relatively thick scintillator in order to increase the detector efficiency at high frequencies where the dependence on Lorentz factor is greatest. (The detectors must still be kept sufficiently
thin so that nuclear interactions, multiple scattering, and bremsstrahlung do not become unacceptably large. Also, in a space instrument, the thickness may be limited in order to keep the total mass acceptably small.)

Figure \ref{ACCESS} shows a schematic of the CSTRD proposed for the ACCESS cosmic ray composition experiment. The instrument consists of five layers of Al honeycomb (effective dimensions $l_1 = 150\  \mu$m, $l_2 = 1$ cm, N = 30 foils in each radiator), each followed by a 2 mm thick CsI(Na) scintillator. The lateral dimensions are 1.6 m $\times$ 1.6 m. The scintillator is divided into strips 1.9 cm wide $\times$ 1.6 m long so that X-rays scattered out of the incident particle beam can be identified. The detector is surrounded on six sides by a set of 1 cm thick CsI scintillator layers. The presence of both horizontal and vertical scintillator layers on all six sides of the detector reflects the broad angular distribution of the scattered radiation and provides a wide field of view for an isotropic flux of incident particles. The choice of detector parameters involves a number of trade-offs. As shown in Fig.\ \ref{comparison}, the efficiency decreases above 200 keV, even though Fig.\ \ref{calculated_spectra} demonstrates that the energy dependence is greatest above 200 keV. Thicker scintillators would increase the detected signal, but then the detector weight and thickness along the particle beam (which translates into interaction probability) would increase unacceptably. The detector parameters chosen reflect a compromise between performance and practicality. Fig.\ \ref{yield_vs_gamma} shows the calculation of expected yield for cosmic ray iron nuclei with the requirement that the particles do not undergo a nuclear interaction as they pass vertically through the detector, and taking into account photoelectron absorption, fluorescence production and escape, and multiple Compton scattering. The set of detector parameters chosen for the ACCESS design results in a calculated signal that increases with energy and saturates at a Lorentz factor of approximately $10^5$.

\section{V.\ Conclusions}

Transition radiation detectors have been used at Lorentz factors up to  $\gamma \sim 10^3 - 10^4$. 
In order to apply the technique at $\gamma$ as high as $10^5$, metal 
radiators can be used with relatively thick foils and large spacings.
The radiation then appears between 0.1 and several MeV energies, the photons are Compton scattered out of the beam,
and scintillators can be used to detect the hard photons. We show that such an approach provides a practical 
method for applying 
the TRD technique to very high energy particle measurements, and describe the differences in the intensity 
and spectrum produced by metal foils and the standard insulating plastic foils or foam.

\begin{acknowledgments}
This work was supported by NASA NAG5-5177 and NASA/Louisiana Board of Regents grant 
NASA/LEQSF-IMP-02. We appreciate numerous conversations with J.\ P.\ Wefel, T.\ G.\ Guzik, and the members of the ACCESS collaboration.
\end{acknowledgments}

\bibliography{tr_in_metal_h}

\end{document}